\documentstyle[12pt]{article}

\newcommand{\newc}{\newcommand}
\newc{\beq}{\begin{equation}}
\newc{\eeq}{\end{equation}}
\newc{\bea}{\begin{eqnarray}}
\newc{\eea}{\end{eqnarray}}
\newc{\kev}{\,\mbox{keV}}
\newc{\gev}{\,\mbox{GeV}}
\newc{\tev}{\,\mbox{TeV}}
\newc{\mev}{\,\mbox{MeV}}
\newc{\ev}{\,\mbox{eV}}
\newc{\gsim}{\lower.7ex\hbox{$\;\stackrel{\textstyle>}{\sim}\;$}}
\newc{\lsim}{\lower.7ex\hbox{$\;\stackrel{\textstyle<}{\sim}\;$}}
\newc{\mz}{m_Z}
\newc{\mpl}{M_{Pl}}
\newc{\obs}{{{\cal O}}}
\newc{\nn}{\nonumber}
\newc{\ie}{{\it i.e.}}
\newc{\eg}{{\it e.g.}}
\newc{\etal}{{\it et al.}}
\newc{\Z}{{\cal Z}}
\newc{\N}{{\cal N}}

\catcode`@=11
\long\def\@caption#1[#2]#3{\par\addcontentsline{\csname
  ext@#1\endcsname}{#1}{\protect\numberline{\csname
  the#1\endcsname}{\ignorespaces #2}}\begingroup
    \small
    \@parboxrestore
    \@makecaption{\csname fnum@#1\endcsname}{\ignorespaces #3}\par
  \endgroup}
\catcode`@=12

\begin{document}
\begin{titlepage}
{
\flushleft{hep-ph//9802314} \hfill FERMILAB-Pub-98/052-T\vspace*{-0.17in}
\flushleft{February 1998} \hfill IASSNS-HEP-98-14\vspace*{-0.17in}
}
{\flushleft ~ \hfill {BA-98-06}}
\vskip 1.0cm
\begin{center}
{\LARGE A Minimality Condition and}\\
\vskip 0.2cm
{\LARGE  Atmospheric Neutrino Oscillations}\footnote{Work supported in part
by Department of Energy Grant Numbers DE-FG02-91ER-40626,
DE-FG02-90ER-40542.} \\
\vskip 0.5cm
{\large\bf Carl H. Albright}$^{a,b,}$\footnote{Email: {\tt albright.fnal.gov,
$^3$babu@ias.edu, $^4$smbarr@bartol.udel.edu}},
{\large\bf K.S. Babu}$^{c,3}$, and
{\large\bf S.M. Barr$^{d,4}$\\
}
\vskip 0.4cm
{\large\sl $^a$Department of Physics, Northern Illinois University, \\
DeKalb, IL, 60115\\}
\vskip 0.4cm
{\large\sl $^b$Fermi National Accelerator Laboratory, P.O. Box 500,\\
Batavia, IL, 60510\\}
\vskip 0.4cm
{\large \sl $^c$School of Natural Sciences, Institute for Advanced Study,\\
Princeton, NJ, 08540\\}
\vskip 0.4cm
{\large \sl $^d$Bartol Research Institute, University of Delaware,\\
Newark, DE, 19716\\}
\end{center}
\vskip .5cm

\baselineskip = 0.25in

\begin{abstract}

A structure is proposed for the mass matrices of the quarks
and leptons that arises in a natural way from the assumption
that the breaking of $SO(10)$ gauge symmetry is achieved by the smallest
possible set of vacuum expectation values. This structure
explains well many features of the observed spectrum of quarks and leptons.
It reproduces the Georgi-Jarlskog mass relations and postdicts the
charm quark mass in reasonable agreement with data.
It also predicts a large mixing angle between $\nu_{\mu}$ and
$\nu_{\tau}$, as suggested by atmospheric neutrino data. The
mixing angles of the electron neutrino are predicted to be small.

\end{abstract}
\end{titlepage}
\setcounter{footnote}{0}
\setcounter{page}{1}
\setcounter{section}{0}
\setcounter{subsection}{0}
\setcounter{subsubsection}{0}


\baselineskip = 0.225in

In this Letter we propose a structure for the quark and lepton
mass matrices that arises naturally in supersymmetric
$SO(10)$ from the
simple assumption that {\it $SO(10)$ is broken to the Standard
Model by the smallest possible set of vacuum expectation
values} (VEVs). This structure reproduces many of the  
features of the known fermion mass spectrum. It also
predicts a large value for the $\nu_{\mu}-\nu_{\tau}$ mixing angle,
as is suggested by the atmospheric neutrino data [1]. Usually
this angle is small (or not predicted) in grand unified
theories, but in the present model its large value has
a simple group-theoretical explanation.

The smallest set of vacuum expectation values that can break $SO(10)$
to the Standard Model consists of one adjoint (${\bf 45}$) and one
spinor (${\bf 16}$) [2]. The spinor plays two necessary roles: it
breaks the rank of the group from 5 to 4, and provides superlarge masses
for the right-handed neutrinos. The adjoint also plays two roles: it
completes the breaking of $SO(10)$ to the Standard Model (SM) group
$SU(3) \times SU(2) \times U(1)$ and produces without fine-tuning 
the ``doublet-triplet splitting" --- that is, gives superlarge
mass to the color-triplet partners of the SM Higgs doublets, while
leaving those doublets light.

Our assumption of minimality requires that there is only {\bf one}
adjoint Higgs. It has recently been shown that this is enough to break
$SO(10)$ with no fine-tuning, while preserving gauge-coupling
unification [3]. Besides its economy, the postulate of having only
one adjoint seems to be desirable in the context of perturbative
heterotic string theory where there are limitations on multiple
adjoints [4]. With only one adjoint,
its VEV is fixed to be in the $B-L$ direction,
as this is required by the Dimopoulos-Wilczek mechanism for
doublet-triplet splitting [5,3]. The superlarge VEV of the spinor
is, of course, also fixed: it must point in the $SU(5)$ singlet direction.
With $SO(10)$ broken to the SM group by only these two definite
VEVs, the possibilities for constructing realistic quark and lepton
masses are quite constrained.  This should be contrasted with other
approaches that generate mass matrix textures in $SO(10)$ utilizing
extended Higgs sector [6].

In ``minimal $SO(10)$" the quark and lepton masses come
from the operators ${\bf 16}_i {\bf 16}_j {\bf 10}_H$, where
$i$ and $j$ are family indices, and subscript $H$ denotes a
Higgs field. This leads to the ``naive $SO(10)$ relations": $N = U
\propto D = L$, with all these matrices being symmetric.
($U$, $D$, $L$, and $N$ denote, respectively,
the mass matrices for the up quarks,
down quarks, and charged leptons, and the Dirac mass matrix of
the neutrinos.)

These relations, as is well known,
lead to bad predictions: $U \propto D$ gives 
vanishing Cabibbo-Kobayashi-Maskawa angles and the relation
$m_c^0/m_t^0 = m_s^0/m_b^0$, which is off by about an order of
magnitude.  (Superscript zero
refers to parameters evaluated at the unification scale.)
One way that $U \propto D$ can be avoided is by the quark and lepton mass
matrices depending on $\langle {\bf 16}_H \rangle$, which breaks
$SO(10)$. ($\langle
{\bf 45}_H \rangle$ does not help here, as up and down quarks have
the same $B-L$.) However, $\langle {\bf 16}_H \rangle$ by itself
leaves $SU(5)$ unbroken, which would still imply the ``naive $SU(5)$ 
relation" $D = L^T$. This contains both the good prediction
$m_b^0 = m_{\tau}^0$, and the bad predictions
$m_s^0 = m_{\mu}^0$, and $m_d^0 = m_e^0$.
Therefore, the quark and lepton mass matrices must also depend
directly or indirectly on
$\langle {\bf 45}_H \rangle$, which is the only $SU(5)$-breaking
VEV. Empirically, one finds the so-called Georgi-Jarlskog
relations [7], $m_s^0 \cong m_{\mu}^0/3$ and $m_d^0 \cong 3 m_e^0$.
Since $\langle {\bf 45}_H \rangle \propto B-L$,
a natural explanation of the Georgi-Jarlskog factors of $3$ and
$1/3$ is possible, as will be shown.

The assumption of minimal VEVs for $SO(10)$-breaking leads naturally,
as will be seen, to the following forms for the
quark and lepton mass matrices at the unification scale
(with the convention that the left-handed fermions
multiply them from the right, and the left-handed antifermions from
the left):

\begin{equation}
\begin{array}{ll}
U^0 = \left( \begin{array}{ccc} 0 & 0 & 0 \\ 0 & 0 & \epsilon/3 \\
0 & - \epsilon/3 & 1 \end{array} \right) m_U, & \;\;\;
N^0 = \left( \begin{array}{ccc} 0 & 0 & 0 \\ 0 & 0 & - \epsilon \\
0 & \epsilon & 1 \end{array} \right) m_U, \\ \\
D^0 = \left( \begin{array}{ccc} 0 & 0 & 0 \\
0 & 0 & \rho + \epsilon/3 \\
0 & -\epsilon/3 & 1 \end{array} \right) m_D, & \;\;\;
L^0 = \left( \begin{array}{ccc} 0 & 0 & 0 \\
0 & 0 & - \epsilon \\ 0 & \rho + \epsilon  & 1
\end{array} \right) m_D.
\end{array}
\end{equation}

These matrices, since they leave $u$, $d$, and $e^-$ massless,
are obviously not the whole story. At the end of this Letter,
we will discuss extending the model to include the first
generation. However,
since $m_e \ll m_\mu$, $m_d \ll m_s$, and $m_u \ll m_c$,
the effects of such first-generation physics should be quite
small on the second and third generation parameters that we wish to fit.
It turns out that with only two parameters,
$\epsilon$ and $\rho$, one can get a good fit for five
quantities that involve the second and third generations:
$m_c/m_t$, $m_s/m_b$, $m_{\mu}/m_{\tau}$, $m_b/m_{\tau}$,
and $V_{cb}$.
(The other mass ratio, $m_b/m_t$ depends on an unknown ratio
of VEVs.)

To give some insight into the structure of the matrices of Eq. (1), and
why they arise naturally from the assumption of minimal VEVs, it
will help to explain how they are built up logically, layer by layer,
from the heaviest generation to the lightest. Because the third generation
is by far the heaviest, and approximately satisfies the $SU(5)$ relation
$m_b^0 = m_{\tau}^0$, we take the first layer to come from the simple 
term ${\bf 16}_3 {\bf 16}_3 {\bf 10}_H$, giving the ``1" entries in Eq. (1).

The second-generation masses, because of the Georgi-Jarlskog
factors, must depend on $\langle {\bf 45}_H \rangle$. The simplest
choice is ${\bf 16}_2 {\bf 16}_3 {\bf 10}_H {\bf 45}_H$. This
gives the ``$\epsilon$" entries in Eq. (1), the
factors of $1/3$ just reflecting the fact that $\langle {\bf 45}_H
\rangle \propto B-L$ and that a quark has
$B-L = 1/3$. It can be shown that $\langle {\bf 45}_H \rangle
\propto B-L$ also implies that this term contributes anti-symmetrically
in flavor. (For this reason the terms 
${\bf 16}_2 {\bf 16}_2 {\bf 10}_H {\bf 45}_H$
and ${\bf 16}_3 {\bf 16}_3 {\bf 10}_H {\bf 45}_H$
would not contribute.) 

The matrices with only the ``1" and ``$\epsilon$" entries, but
without ``$\rho$", would still not be realistic: the matrices $U$ and $D$ 
would be proportional, giving $V_{cb}= 0$ and $m_c^0/m_t^0 = m_s^0/m_b^0$,
and the Georgi-Jarlskog factor would be 9 instead of 3.
(The ``see-saw" formula would give $m_{\mu}^0/m_{\tau}^0 \cong 
\epsilon^2$, and $m_s^0/m_b^0
\cong \epsilon^2/9$.)

It turns out that all three of these unrealistic features are cured in a single
stroke by introducing a third layer that involves $\langle {\bf 16}_H \rangle$.
The simplest term, group theoretically, that can be written down
is of the form ${\bf 16}_2 {\bf 16}_3 {\bf 16}_H {\bf 16}'_H$.
${\bf 16}'_H$ is some
spinor Higgs, distinct from ${\bf 16}_H$, which breaks
the electroweak symmetry but does {\it not}
participate in the breaking of $SO(10)$ down to the Standard Model group [8]. 
(That is, the components that get VEVs are ${\bf 1}({\bf 16}_H)$, and
$\overline{{\bf 5}}({\bf 16}'_H)$, where ${\bf p}({\bf q})$ denotes a
${\bf p}$ of $SU(5)$ contained in a ${\bf q}$ of $SO(10)$.)

This term arises most naturally from ``integrating out" ${\bf 10}$'s
of $SO(10)$, as shown in  Figure 1. The resulting operator is
$\overline{{\bf 5}}({\bf 16}_2) {\bf 10}({\bf 16}_3)
\langle \overline{{\bf 5}}({\bf 16}'_H) \rangle \langle
{\bf 1}({\bf 16}_H) \rangle$. Note that this contributes to $L$
and $D$, but not to $U$ and $N$, and that it lopsidedly contributes
to $D_{23}$ and $L_{32}$ but not to $D_{32}$ and $L_{23}$.
This is the origin of the ``$\rho$" entries in Eq. (1).
This lopsidedness, which has a group-theoretical origin,
explains, as will be seen, why the 2-3 mixing is small for
the quarks ($V_{cb} \ll 1$) but large for the leptons ($\sin^2 2
\theta_{\mu \tau} \sim 1$).

There can be a relative phase, which we will call $\alpha$, between
the parameters $\epsilon$ and $\rho$. As is apparent from Eq. (1),
this phase only enters at order $\epsilon/\rho$, which will presently be
seen to be a small parameter. (Henceforth the symbols
$\rho$ and $\epsilon$ will denote $|\rho|$ and $|\epsilon|$, and
the phase will appear explicitly as $\alpha$.) Diagonalizing the matrices
in Eq. (1), one finds:

\begin{equation}
\begin{array}{l}
m_b^0/m_{\tau}^0 \cong 1 - \frac{2}{3} \frac{\rho}{\rho^2 + 1}
(\epsilon \cos \alpha), \\ \\
m_{\mu}^0/m_{\tau}^0 \cong \epsilon \frac{\rho}{\rho^2 + 1}
\left( 1 - \frac{\rho^2 - 1}{\rho (\rho^2 + 1)}(\epsilon \cos \alpha)
\right), \\ \\
m_s^0/m_b^0 \cong \frac{1}{3} \epsilon \frac{\rho}{\rho^2 + 1}
\left( 1 - \frac{1}{3}
\frac{\rho^2 - 1}{\rho (\rho^2 + 1)}(\epsilon \cos \alpha)
\right), \\ \\
m_c^0/m_t^0 \cong \epsilon^2 /9, \\ \\
V_{cb}^0 \cong \frac{1}{3} \epsilon \frac{\rho^2}{\rho^2 + 1}
\left( 1 + \frac{2}{3} \frac{1}{\rho (\rho^2 + 1)} (\epsilon \cos \alpha)
\right).
\end{array}
\end{equation}

\noindent
In these expressions terms that are down by order $O(\epsilon^2)$
have been dropped. (They affect the results at the fraction of a
percent level.)
Because $\epsilon$ is a small parameter, the following
features of the observed masses and mixings
have been reproduced by the model: the approximate
equality of $m_b^0$
and $m_{\tau}^0$; the fact that $V_{cb}^0$, $m_{\mu}^0/m_{\tau}^0$, and
$m_s^0/m_b^0$ are all comparable, because $O(\epsilon)$, while $m_c^0/m_t^0$ is
very much smaller, because $O(\epsilon^2)$; and the fact
that $m_{\mu}^0/m_{\tau}^0$ is about $3$ times $m_s^0/m_b^0$
(one of the the Georgi-Jarlskog relations).
Also explained is the hierarchy among generations, which arises
from the smallness
of $\epsilon$ and from the rank-2 nature of the matrices.

Since there are five observables in terms of the two parameters
$\epsilon$ and $\rho$ in Eq. (2), the model predicts three relations
among charged fermions.
To study them we use the following input parameters:
$m_{\mu} = 105.66$ MeV, $m_{\tau} = 1.777$ GeV, $m_s(1 \; {\rm GeV}) =
(180 \pm 50)$ MeV, $m_b(m_b) = (4.26 \pm 0.11)$ GeV, $m_c(m_c) =
(1.27 \pm 0.1)$ GeV [9], $M_t = 174.1 \pm 5.4$ GeV
and $V_{cb} = 0.0395 \pm 0.0017$ [10].  The value of $M_t$ quoted
above corresponds to the running masses $m_t(m_t) = 165 \pm 5~GeV$.

To fit the data, various renormalization factors are needed.
The factors, that will be denoted by $\eta_i$, 
that run the masses from the low scales up to the
supersymmetry scale, $M_{SUSY}$ (taken to be at $m_t$)
are computed using
3-loop QCD and 1-loop QED or electroweak renormalization group
equations (RGE), with inputs $\alpha_s (M_Z) = 0.118$,
$\alpha (M_Z) = 1/127.9$ and $\sin^2 \theta_W (M_Z) = 0.2315$.
The relevant RGE can be found for instance in [11]. The results are
($\eta_{\mu}, \eta_\tau, \eta_s, \eta_b,\eta_c, \eta_t) =
(0.982, 0.984, 0.426, 0.654, 0.473,1.0)$.

The renormalization factors from $M_{SUSY}$ up to the unification scale,
$M_G$, are calculated using the 2-loop MSSM beta functions for
all parameters [11], with $M_G = 2 \times 10^{16}$ GeV, and
all SUSY thresholds taken to be at $M_{SUSY}$.  These factors also
depend on the value of tan$\beta$, which is allowed {\it a priori}
(by the perturbativity of the Yukawa couplings up to $M_G$)
to be anywhere in the range $1.5 \le {\rm tan}\beta
\le 65$. However, as will be seen, within our scheme the fits constrain
tan$\beta$ to be between 10 and 40. Results will be presented for a 
``central" value of 30, and where significant the dependence on tan$\beta$
will be discussed. (In this model, since
the light doublet, $H'$ is a linear combination of
$\overline{{\bf 5}}({\bf 10})$ and $\overline{{\bf 5}}({\bf 16}')$,
$\tan \beta \ne m_t/m_b$.
It is also not expected to be very small, since the same Yukawa
coupling contributes to both the top and bottom quark masses.)
The running factors for tan$\beta=30$ are
$(\eta_{\mu/ \tau}, \eta_{s/b}, \eta_{c/t},\eta_{b/\tau}, \eta_{cb})  =
(0.956, 0.840, 0.691,0.514,0.873)$, where $\eta_{i/j}
\equiv (m_i^0/m_j^0)/(m_i/m_j)_{M_{SUSY}}$, and $\eta_{cb}
\equiv V_{cb}^0/(V_{cb})_{M_{SUSY}}$.

Aside from the running of the couplings described by the $\eta$'s,
there are finite corrections [12] to $m_s, m_b$ and $V_{cb}$ from
gluino and chargino loops, which are proportional to tan$\beta$ and
thus sizable for moderate to large tan$\beta$.
These will be denoted by the factors
$(1 + \Delta_{s})$, $(1 + \Delta_b)$, and $(1 + \Delta_{cb})$,
which depend on the supersymmetric spectrum:
$\Delta_b \simeq {\rm tan}\beta\{ {2 \alpha_3 \over 3 \pi}
{\mu M_{\tilde{g}} \over
m_{\tilde{b}_L}^2-m_{\tilde{b}_R}^2}[f(m_{\tilde{b}_L}^2/M_{\tilde{g}}^2) -
f(m_{\tilde{b}_R}^2/M_{\tilde{g}}^2)]
+ {\lambda_t^2 \over 16 \pi^2} {\mu A_t \over m^2_{\tilde{t}_L}
-m^2_{\tilde{t}_R} } [f(m^2_{\tilde{t}_L}/\mu^2) - f(m^2_{\tilde{
t}_R}/\mu^2)]\}$, where $f(x) \equiv {\rm ln}x/(1-x)$.
$\Delta_s$ is given by the same expression
but without the chargino contribution (the second term) and with 
$\tilde{b} \rightarrow \tilde{s}$. $\Delta_{cb} =
-\Delta_b^{\rm chargino}$.  One sees that even for tan$\beta \approx
10$, these corrections are of order 10\%.
The analogous corrections to $m_{\mu}$ and $m_{\tau}$ arise only
from Bino loops, while those to $m_c$ and $m_t$ lack the tan$\beta$
enhancement, and so these are all negligible.

To fit for $\rho$ and $\epsilon$ it is convenient to use the 
second and fifth relations of Eq. (2), since there is very little 
experimental uncertainty in $m_{\mu}$, $m_{\tau}$, and $V_{cb}$.
This gives $\rho = [3 V_{cb}/(m_{\mu}/m_{\tau})] 
(\frac{\eta_{\tau} \eta_{cb}}
{\eta_{\mu} \eta_{\mu/\tau}} )$ $( 1
- \frac{\epsilon \cos \alpha}{3}
\frac{3 \rho^2 - 1}{ \rho (\rho^2 + 1)} ) (1 - \Delta_{cb})$,
and $\epsilon = [\frac{\rho^2 +1}{\rho}
(m_{\mu}/m_{\tau})] ( \frac{\eta_{\mu} \eta_{\mu/\tau}}{\eta_{\tau}})$
$( 1 + \epsilon \cos \alpha
\frac{\rho^2 - 1}{\rho (\rho^2 + 1)})$.
One finds, for $\cos \alpha = 1$, that

\begin{equation}
\rho = 1.73 \; (1 - \Delta_{cb}), \;\;\;\;\;
\epsilon = 0.136 \; (1 - 0.5 \Delta_{cb}).
\end{equation}

\noindent
The dependence on
$\cos \alpha$, arising only at order
$\epsilon/\rho$, is rather weak: for $\cos \alpha = -1$,
$\rho = 1.92 \; (1 - \Delta_{cb})$ and
$\epsilon = 0.134 \; (1 - 0.5 \Delta_{cb})$. The dependence on
$\tan \beta$, because it is only through
the renormalization factors, is also fairly weak for
$10 \leq \tan \beta \leq 40$. For example,
increasing $\tan \beta$ to $40$ increases $\rho$ by $0.7 \%$
and decreases $\epsilon$ by $3 \%$.  Similarly, changing
$M_{SUSY}$ from $m_t$ to $500~GeV$ only
increases $\rho$ by 3\% and increases $\epsilon$ by 2\%.
Henceforth, all results will
be stated for $\tan \beta = 30$, $M_{SUSY} = m_t$,
and $\cos \alpha = 1$.  Whenever results are very sensitive
to these parameters, the dependence on them will be explicitly
discussed.

Now that $\rho$ and $\epsilon$ have been determined from $V_{cb}$
and $m_{\mu}/m_{\tau}$, there are four other quantities that can be
predicted, namely $m_b$, $m_s$, $m_c$, and $\sin^2 2 \theta_{\mu \tau}$.

\vspace{0.2cm}

\noindent
{\bf (i) $m_b$ prediction:}

The first relation of Eq. (2) implies
$m_b(m_b) = m_{\tau}(m_\tau) (\frac{\eta_{\tau}}{\eta_b \eta_{b/\tau}})
( 1 - \frac{2}{3} \frac{\rho}{\rho^2 + 1} \epsilon
\cos \alpha )$ $(1 + \Delta_b)$. For $\cos \alpha = 1$, this gives
$m_b(m_b) = 5.0 \; (1 + \Delta_b)$ GeV. Comparing this with
the experimental value $4.26 \pm 0.11$ GeV, one sees that $\Delta_b
\cong - 0.15$.  This is quite a reasonable value
if tan$\beta \approx 30$.  (With supergravity boundary conditions
and a generic sparticle spectrum, the
gluino loops contribute $\sim \pm 0.2$ to $\Delta_b$, 
while the charginos contribute roughly a quarter as much and with the
opposite sign [13].  We shall keep these numbers as a rough guide
to estimate the corrections.)
It should be noted
that if tan$\beta$ is close to $1.6$ or near $60$, $m_b(m_b)$ will be
in the acceptable range
even with $\Delta_b = 0$.  However, these extreme values of
tan$\beta$ lead to wrong predictions of the charm mass ($m_c(m_c) \simeq
1.57~GeV$ when tan$\beta\simeq 1.6$)
and are thus disfavored within the model.
An interesting consequence is that the model
predicts the sign of $\mu$ (and $A_t$) to be such that it decreases the
$b$--quark mass through the gluino and chargino graphs.

\vspace{0.2cm}

\noindent
{\bf (ii) $m_s$ prediction:}

The first and third relation
of Eq.(2) yield $m_s(1~GeV) = m_{\tau}(m_\tau) \frac{1}{3} \epsilon
\frac{\rho}{\rho^2 + 1}(\frac{\eta_{\tau}}{\eta_s \eta_{s/b}
\eta_{b/\tau}})$  $( 1 - \frac{1}{3} \epsilon \cos \alpha
\frac{3 \rho^2 - 1}{\rho (\rho^2 + 1)})( 1 + \Delta_s)$.
For $\cos \alpha = 1$ this gives $m_s(1 \; GeV) = 176 \; (1 + \Delta_s)$
MeV.  Taking $\Delta_s \simeq \Delta_b \cong -0.15$,
which is justified if the
gluino contribution dominates and  $\tilde{s}$ and $\tilde{b}$
are nearly degenerate, we find
$m_s(1~GeV)= 150$ MeV, in excellent agreement with
the experimental value of $180 \pm 50$ MeV.

\vspace{0.2cm}

\noindent
{\bf (iii) $m_c$ prediction:}

The fourth relation of
Eq. (2) implies $m_c(m_c) = m_t(m_t) \frac{1}{9} \epsilon^2 (\frac{\eta_t}
{\eta_c \eta_{c/t}})$. For $\cos \alpha = 1$,
this gives $m_c = (1.05 \pm 0.11)
(1 - \Delta_{cb})$ GeV. The error reflects the $1\sigma$ uncertainties in
the experimental values of $m_t$, $\alpha_s$ ($=0.118 \pm 0.004$),
and $V_{cb}$.
(These lead, respectively, to
$6.5 \%$, $7 \%$, and $4 \%$ uncertainties for $m_c(m_c)$. It should also
be noted that changing $\tan \beta$ by $\pm 10$ changes the
$m_c$ prediction by $\mp 4 \%$, changing $M_{SUSY}$ to
$500$ GeV has less than a $2 \%$ effect, and changing $\cos \alpha$
to $0$ reduces $m_c$ by $3 \%$.) Since
$\Delta_{cb} \simeq - \Delta_b|^{\rm chargino}$, it is reasonable to
take $\Delta_{cb} \simeq-0.05$, using the supergravity--like spectrum
as a guideline.  This gives $m_c = 1.10 \pm 0.11$ GeV.
This is in quite reasonable agreement with the experimental
value $m_c(m_c) = (1.27 \pm 0.1)$ GeV.  It is interesting that
the sign of the correction term $\Delta_{cb}$ suggested by the
supergravity spectrum is such that it improves the agreement of $m_c(m_c)$
with the experimental value.

It should also be emphasized that, whereas the predictions for $m_b$
and $m_s$ were, in a sense, group-theoretically built into the
forms given in Eq. (1), it could not be known in advance that the
prediction for $m_c$ would come close.

\vspace{0.2cm}

\noindent
{\bf (iv) $\sin^2 2 \theta_{\mu \tau}$ prediction:}

The neutrino-mixing matrix $U_{\nu}$ is defined by
$\nu_f = \sum_m (U_{\nu})_{fm} \nu_m$, where, $\nu_f$ and $\nu_m$ are the
flavor and mass eigenstates, respectively. $f = e, \mu, \tau$, and
$m = 1,2,3$. $U_{\nu} = U^{(L) \dag} U^{(N)}$,
where $U^{(L)}$ and $U^{(N)}$ are the unitary transformations
of the left-handed fermions required to diagonalize, respectively,
$L$ and $M_{\nu} = - N^T M_R^{-1} N$. ($M_R$ is the superheavy Majorana
mass matrix of the right-handed neutrinos.)

The crucial point, easily seen from an inspection of Eq. (1), is that
a large rotation in the 2-3 plane will be required to diagonalize
the charged-lepton mass matrix, $L$. Calling this rotation angle
$\theta_{23}^{(L)}$, one has that $\tan \theta_{23}^{(L)} \cong
\left| L_{32}/L_{33}\right| \cong \rho + \epsilon \cos \alpha$.
The actual $\nu_{\mu}-\nu_{\tau}$ mixing angle is the difference
between $\theta_{23}^{(L)}$ and the corresponding rotation angle,
$\theta_{23}^{(N)}$, required to diagonalize $M_{\nu}$.

It might appear that one can know nothing about $M_{\nu}$, and therefore
about $\theta_{23}^{(N)}$, without knowing the precise form of
$M_R$. However, this is not the case. From Eq. (1) it is apparent
that in the limit $\epsilon \longrightarrow 0$ both $N$ and
$M_{\nu} = - N^T M_R^{-1} N$ are proportional
to ${\rm diag}(0,0,1)$, so that $\theta_{23}^{(N)} \longrightarrow 0$.
Thus, formally, $\theta_{23}^{(N)} = O(\epsilon)$. If $M_R^{-1}$ is
parametrized by $(M_R^{-1})_{22} = (M_R^{-1})_{33} Y/\epsilon^2$, and
$(M_R^{-1})_{23} = (M_R^{-1})_{32} = (M_R^{-1})_{33} X/\epsilon$,
one finds (ignoring
the first generation) that $\tan 2 \theta_{23}^{(N)}
\cong 2 \epsilon \left|(1 - X)/
(1 - 2 X + Y)\right|$.
Unless $X$ and $Y$ are fine-tuned, this is indeed of order
$\epsilon$. Let $\kappa$ be defined by
${\rm Re}(U^{(N)}_{23}) = \kappa \epsilon \cos \alpha$, in a
phase convention where $U^{(L)}_{23}$ is real.
If it is required that $m_{\nu_{\mu}}/m_{\nu_{\tau}} \approx 0.05$,
as suggested by the atmospheric and solar neutrino data, then
$\left| \kappa \right| \stackrel{_<}{_\sim} 2$.
The $\mu-\tau$ mixing angle at the unification scale is then
given by

\begin{equation}
\tan \theta_{\mu \tau} = \frac{\rho + (1 - \kappa) \epsilon \cos \alpha}
{1 + \kappa \rho \epsilon \cos \alpha}.
\end{equation}

\noindent
The one-loop renormalization group equation for this quantity [14] has
the simple form
$d (\ln \tan \theta_{\mu \tau})/d (\ln \mu) = - h_{\tau}^2/16 \pi^2$.
Integrating yields the result that (for $\tan \beta = 30$)
$\tan \theta_{\mu \tau}$
is increased by a factor of $1.03$ in running down to the weak
scale from the unification scale.

Unlike the quark masses, the $\nu_{\mu}-\nu_{\tau}$ mixing angle is very
sensitive to $\cos \alpha$, and therefore $\sin^2 2 \theta_{\mu \tau}$
can be in a large range, from 1 down to about $1/4$. Values
$> 0.7$ obtain for most of the parameter range. For example,
if $\cos \alpha = 0$, Eq. (4) simplifies to
$\tan \theta_{\mu \tau} = \rho$, giving
$\sin^2 2 \theta_{\mu \tau} = 0.78$, independent of $\kappa$.
If $\kappa = 0$, then $\sin^2 2 \theta_{\mu \tau} > 0.7$ for all
$\cos \alpha$. $\sin^2 2 \theta_{\mu \tau}$ reaches $1$ for
$\cos \alpha = 1$ and $\kappa = 2$, and reaches $\approx 1/4$ for
$\cos \alpha = 1$ and $\kappa = -2$.
(For recent
attempts to generate large $(\nu_\mu-\nu_\tau)$ mixing in
other ways see Ref. [15].)

There is not a unique way to extend this model to include
the first generation. A simple possibility that gives
a reasonable fit to the first-generation masses and mixings
is to add $(12)$ and $(21)$ entries symmetrically to all the mass matrices.
This would give several new predictions: (i) ${m_d^0
\over m_e^0} =
3(1+{2 \over 3 \rho} \epsilon {\rm cos}\alpha)$ (one of the
Georgi--Jarlskog relations), (ii) $|V_{us}^0| = |\sqrt{m_d^0 \over m_s^0}
{1 \over (\rho^2+1)^{1/4} } - \sqrt{m_u^0 \over m_c^0}e^{i\phi}|$,
(iii) $|V_{ub}^0| \simeq |\sqrt{m_d^0 \over m_s^0} {m_s^0 \over m_b^0}
{\rho \over (\rho^2+1)^{1/4} } - \sqrt{m_u^0 \over m_c^0} e^{i\phi}
(\sqrt{m_c^0 \over m_t^0} - {m_s^0 \over m_b^0}{1\over\rho})|$.
If the phase parameter $\phi$ is near $\pi$, acceptable
$|V_{us}|$ and $|V_{ub}|$ result.  The leptonic mixing angles involving
the electron are given by $|(U_{\nu})^0_{e2}| \simeq 
|\sqrt{m_e^0 \over m^0_\mu}
(\rho^2+1)^{1/4} + O(\epsilon)|$, and $|(U_{\nu})^0_{e3}| \simeq
|\sqrt{m_e^0 \over m_\mu^0}
{m_\mu^0 \over m_\tau^0} {(\rho^2+1)^{3/4} \over \rho} + O(\epsilon^2)|$
where the $O(\epsilon)$ and $O(\epsilon^2)$ terms represent corrections
from the neutrino sector.  Since these mixing angles are both small,
their precise values are sensitive to the structure of $M_R$.
These values are consistent with the small angle MSW oscillations
for the solar neutrinos.

The model presented here can be tested in future
experiments in several ways.
(i) The prediction of tan$\beta = 10-40$ can be tested once
supersymmetric particles are discovered.  (ii) The spectrum of
the sparticles is predicted to be such that the gluino and the chargino
corrections to $m_b$ decrease its value by about 15 \%.
(iii) More precise determinations of $m_t, \alpha_3(M_Z)$ and
$V_{cb}$ and information about the sparticle spectrum 
will sharpen the model's prediction of $m_c(m_c)$.
(iv) Large angle $(\nu_\mu-\nu_\tau$) oscillations should be seen
in long baseline experiments, but not in the
ongoing accelerator experiments.  The interpretation of the
atmospheric neutrino anomaly in terms of ($\nu_\mu-\nu_\tau)$ oscillations
should be confirmed.
(v) There are also specific predictions
in the model for proton decay branching ratios [16] and rare decays such
as $\mu \rightarrow e\gamma$ [17].

In this Letter we have studied a simple form for the mass matrices
that is motivated by general group-theoretical considerations,
without examining a particular underlying unified model
in great detail. That has been done in [18], where it is found
that fermion mass matrices of the type discussed here can
arise in realistic models.

\section*{References}

\begin{enumerate}

\item K.S. Hirata {\it et al.}, {\it Phys. Lett.} {\bf 205B},
416 (1988); K.S. Hirata {\it et al.}, {\it Phys. Lett.} {\bf 280B},
146 (1992); Y. Fukuda {\it et al.}, {\it Phys. Lett.} {\bf 335B},
237 (1994); D. Caspar {\it et al.}, {\it Phys. Rev. Lett.}
{\bf 66}, 2561 (1991); R. Becker-Szendy {\it et al.},
{\it Phys. Rev.} {\bf D46}, 3720 (1992); {\it Nucl. Phys.}
{\bf B} (Proc. Suppl.) {\bf 38}, 331 (1995); T. Kafka, {\it Nucl. Phys.}
{\bf B} (Proc. Suppl.) {\bf 35}, 427 (1994); M. Goodman, {\it ibid.}
{\bf 38}, 337 (1995); W.W.M. Allison {\it et al.}, {\it Phys.
Lett.} {\bf 391B}, 491 (1997).
\item In order to break the gauge symmetry at the
unification scale while preserving supersymmetry, it
is generally necessary in $SO(10)$ that
for every ${\bf 16}$ that has a superlarge VEV, there be a
corresponding $\overline{{\bf 16}}$ with an equal VEV.
\item S.M. Barr and S. Raby, {\it Phys. Rev. Lett.} {\bf 79}, 4748 (1997).
\item K. Dienes, {\it Nucl. Phys.} {\bf B488}, 141 (1997); K. Dienes and
J. March-Russell, {\it Nucl. Phys.} {\bf B479}, 113 (1996);  Z. Kakushadze,
G. Shiu and S.H. Tye, hep-th/9710149; J. Erler, {\it Nucl. Phys.} {\bf B475},
597 (1996).
\item S. Dimopoulos and F. Wilczek, Report No. NSF-ITP-82-07 (1981),
in {\it The unity of fundamental interactions}, Proceedings of the 19th
Course of the International School of Subnuclear Physics, Erice, Italy,
1981, ed. A. Zichichi (Plenum Press, New York, 1983);
K.S. Babu and S.M. Barr, {\it Phys. Rev.} {\bf D48}, 5354 (1993).
\item G. Anderson et al, {\it Phys. Rev.} {\bf D49}, 3660 (1994);
A. Kusenko and R. Shrock, {\it Phys. Rev.} {\bf D49}, 4962 (1994);
K.S. Babu and R.N. Mohapatra, {\it Phys. Rev. Lett.} {\bf 74}, 2418 (1995);
L. Hall and S. Raby, {\it Phys. Rev.} {\bf D51}, 6524 (1995);
K.S. Babu and S.M. Barr, {\it Phys. Rev. Lett.} {\bf 75}, 2088 (1995);
C. Albright and S. Nandi, {\it Phys. Rev. Lett.} {\bf 73}, 930 (1994) and
{\it Phys. Rev.} {\bf D 53}, 2699 (1996);
Z. Berezhiani, {\it Phys. Lett.} {\bf B409}, 220 (1997);
K.C. Chou and Y.L. Yu, hep-ph/9708201.
\item H. Georgi and C. Jarlskog, {\it Phys. Lett.} {\bf 86B},
297 (1979).
\item If ${\bf 16}_H$ and ${\bf 16}'_H$ were not distinct,
the operator would be ${\bf 16}_2 {\bf 16}_3 ({\bf 16}_H)^2$. This would
lead to flavor-symmetric contributions to the mass matrices, which would
yield unrealistic predictions.
\item J. Gasser and H. Leutwyler, {\it Phys. Rept.} {\bf 87}, 77 (1982).
\item {\it Table of Partilce Properties},
Particle Data Group, 1998 [http://pdg.lbl.gov/].
\item H. Arason {\it et al.}, {\it Phys. Rev.} {\bf D46}, 3495 (1992);
V. Barger, M. Berger and P. Ohmann, {\it Phys. Rev.} {\bf D47}, 1093
(1993) and
{\it Phys. Rev.} {\bf D47}, 2038 (1993).
\item L. Hall, R. Rattazzi and U. Sarid,  {\it Phys. Rev.} {\bf D50},7048
(1994).
\item T. Blazek, M. Carena, S. Raby, C.E.M. Wagner,
{\it Phys. Rev.} {\bf D56}, 6919 (1997).
\item K.S. Babu, C.N. Leung, and J. Pantaleone, {\it Phys.
Lett.} {\bf 319B}, 191 (1993).
\item B. Brahmachari and R.N. Mohapatra, hep-ph/9710371;
J. Sato and T. Yanagida, hep-ph/9710516;
M. Drees, S. Pakvasa, X. Tata, and T. ter Veldhuis, hep-ph/9712392;
M. Bando, T. Kugo, and K. Yoshioka hep-ph/9710417.
\item K.S. Babu and S.M. Barr, Phys. Lett. {\bf B381}, 137 (1996);
P. Nath, Phys. Rev. Lett. {\bf 76}, 2218 (1996);
K.S. Babu, J.C. Pati and F. Wilczek, hep-ph/9712307.
\item R. Barbieri, L. Hall and A. Strumia, {\it Nucl. Phys.} {\bf B445},
219 (1995);
M. Gomez, and H. Goldberg, {\it Phys. Rev.} {\bf D53}, 5244 (1996).
\item C. Albright and S.M. Barr, hep-ph/9712488.

\end{enumerate}

\section*{Figure Caption}

\noindent
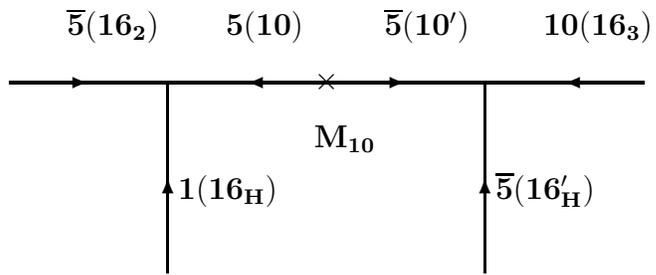
{\bf Figure 1:} A diagram that shows how vectors of fermions may
be integrated out to produce the ``$\rho$" terms
in the mass matrices in Eq. (1). For group-theoretical reasons these
produce lopsided contributions to the charged-lepton and down-quark
mass matrices, that explain why $V_{cb}$ is small while $\sin^2 2
\theta_{\mu \tau}$ is large.

\begin{picture}(360,216)
\thicklines
\put(60,144){\vector(1,0){30}}
\put(90,144){\line(1,0){30}}
\put(120,144){\line(1,0){30}}
\put(180,144){\vector(-1,0){30}}
\put(180,144){\vector(1,0){30}}
\put(210,144){\line(1,0){30}}
\put(240,144){\line(1,0){30}}
\put(300,144){\vector(-1,0){30}}
\put(120,72){\vector(0,1){36}}
\put(120,108){\line(0,1){36}}
\put(175.5,141.5){$\times$}
\put(240,72){\vector(0,1){36}}
\put(240,108){\line(0,1){36}}
\put(82,162){${\bf \overline{5}(16_2)}$}
\put(142,162){${\bf 5(10)}$}
\put(202,162){${\bf \overline{5}(10')}$}
\put(262,162){${\bf 10(16_3)}$}
\put(124,100){${\bf 1(16_H)}$}
\put(175.5,119.5){${\bf M_{10}}$}
\put(244,100){${\bf \overline{5}(16'_H)}$}
\put(162,36){{\bf Fig. 1}}
\end{picture}

\end{document}